\documentclass[aps,twocolumn,showpacs,superscriptaddress]{revtex4}

\usepackage{graphicx}
\usepackage{amsmath}

\DeclareMathOperator{\re}{Re}
\DeclareMathOperator{\im}{Im}

\begin{document}

\title{Synchronization of active mechanical oscillators by an inertial load}
\author{Andrej Vilfan}
\email[]{Andrej.Vilfan@ijs.si}
\affiliation{Cavendish Laboratory, Madingley Road, Cambridge CB3 0HE, UK}
\affiliation{J. Stefan Instutute, Jamova 39, 1000 Ljubljana, Slovenia}
\author{Thomas Duke}
\email[]{td18@cam.ac.uk}
\affiliation{Cavendish Laboratory, Madingley Road, Cambridge CB3 0HE, UK}
\email{andrej.vilfan@ijs.si}
\date{\today}

\newcommand\I{{\rm i}}

\begin{abstract}
  Motivated by the operation of myogenic (self-oscillatory) insect
  flight muscle, we study a model consisting of a large number of
  identical oscillatory contractile elements joined in a chain, whose
  end is attached to a damped mass-spring oscillator. When the
  inertial load is small, the serial coupling favors an
  antisynchronous state in which the extension of one oscillator is
  compensated by the contraction of another, in order to preserve the
  total length. However, a sufficiently massive load can sychronize
  the oscillators and can even induce oscillation in situations where
  isolated elements would be stable. The system has a complex phase
  diagram displaying quiescent, synchronous and antisynchrononous
  phases, as well as an unsual asynchronous phase in which the total
  length of the chain oscillates at a different frequency from the
  individual active elements.
\end{abstract}

\maketitle

The origin of movement in biological systems can frequently be traced
to molecular motors --- specialized proteins that convert chemical
energy to mechanical work. Kinesin and dynein, which travel along
microtubules, and myosin, which pulls on actin filaments, are typical
examples \cite{alberts,Howard_book}.  In many cases, such as muscle
contraction or intracellular transport, molecular motors generate
uni-directional motion. But there are also a number of physiological
systems which incorporate motor proteins that display oscillatory
dynamics.  These include eucaryotic flagella and cilia whose
undulation is driven by dynein molecules, and the flight muscles of
many insects which contract rythmically at a frequency that is out of
step with the excitory neural impulses
\cite{note_myogenic,Machin.Pringle1959,Machin.Pringle1960,Pringle1978,Dickinson97,Josephson.Stokes2000}.
Apparently these systems are self-oscillatory, and the dynamical
instability that leads to vibration is directly generated by the
action of the motor proteins; it is usually attributed to delayed
stretch activation, which can be caused by a variety of different
microscopic mechanisms (reviewed in Refs.~\cite{Pringle1978} \&
\cite{Dickinson97}). Oscillations have also been observed in the
sarcomeres of skeletal muscle in non-physiological conditions
\cite{Okamura.Ishiwata1988,Anazawa.Ishiwata1992,yasuda96,Fujita.Ishiwata1998}
and in experiments that probe the interaction between individual
dynein molecules and microtubules \cite{shingoyi98}. Even normal
muscle fibers can display a damped oscillatory response to sudden
changes in load \cite{Edman.Curtin2001}.

Theoretical analysis has demonstrated that a single filament
interacting with an ensemble of motors can have an anomalous
force-velocity relation \cite{prost95,vilfan99}, whereby two different
sliding speeds, one positive and one negative, can occur at a given
load. Experimental confirmation of this phenomenon has been obtained
in gliding motility assays for both actin- \cite{riveline98} and
microtubule-based systems \cite{Endow.Higuchi2000}. In such a
situation the motors can collectively generate oscillations when the
filament is connected in series with an elastic element
\cite{juelicher97,thomas98,Juelicher2001}, because the solution for
the sliding speed then switches periodically between the two stable
branches.  However, direct application of this model to muscle fibers,
composed of hundreds of sarcomeres (contractile units) in series,
omits a crucial point: the oscillations will only be macroscopically
observable if there is at least some degree of synchrony between the
oscillations of individual sarcomeres. How might this occur? One
possibility is that the activity of myosin motors in different
sarcomeres is coordinated by some chemical signaling. An alternative
suggestion is that torsion of actin filaments is involved
\cite{yasuda96}. In this Letter, we propose a mechanism of
synchronization that does not rely on any such specific molecular
process. We investigate the dynamics of a chain of active mechanical
elements each of which, when isolated, can undergo a dynamical
instability from a quiescent (stable) to an oscillatory (unstable)
regime. In the case where each element is individually stable, we show
that the entire chain can be set into synchronized vibration by the
application of a sufficiently massive inertial load at its end. In the
alternative case where each contractile element is unstable, we
demonstrate the existence of a variety of dynamical regimes, including
an unusual asynchronous state in which individual elements oscillate
at a faster frequency than the mass to which they are connected.

\begin{figure}[t]
  \begin{center}
    \includegraphics{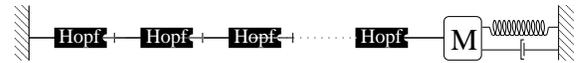}
  \end{center}
  \caption{Model system: A chain of serially coupled active
    Hopf oscillators is attached to a damped inertial oscillator.}
  \label{fig_model}
\end{figure}

The model system that we investigate is shown in Fig.~\ref{fig_model}.
Each contractile element in the chain is an active mechanical system
whose displacement $x_i$ is coupled nonlinearly to an internal
variable $y_i$ (which might, for example, be the fraction of bound
motors, or the concentration of a regulator). We suppose that an
element may either stay still or generate self-sustained oscillations,
depending on the value of a control parameter. In the vicinity of the
critical point where the quiescent state becomes unstable, the
dynamics may be described by the canonical equation for a Hopf
bifurcation. Writing the complex variable $z_i = x_i + {\rm i} y_i$,
we have
\begin{equation}
  \label{eq_zdot}
\dot z_i=(\I \omega + \epsilon) z_i -B\left| z_i \right|^2 z_i + F/\zeta \quad,
\end{equation}
where $\epsilon$ is the control parameter and $\omega$ is the
characteristic frequency of each active oscillator. $F$ is the force
acting on each of the oscillators in the chain, due to the coupled
mass-spring system and is determined by its equation of motion:
\begin{equation}
  \label{eq_f}
  F=-M \sum_{i=1}^{N} \re \left( \ddot z_i +\Omega^2 z_i + \gamma 
  \dot z_i \right) \quad,
\end{equation}
where $M$ is the mass of the load, $\gamma$ is a measure of the
damping and $\Omega$ is the natural frequency of the mass-spring
system.  In order to proceed further with the analysis and obtain
numerical solutions, we replace Eq.~(\ref{eq_f}) with a first-order
differential equation by inserting $\ddot z_i$ from the first
derivative of Eq.~(\ref{eq_zdot})
\begin{multline}
  \label{eq_fdot}
  \dot F=-\frac \zeta {N M} F +\frac \zeta N 
  \sum_{i=1}^{N}\Bigl( -\epsilon \re \dot
   z_i +\omega \im \dot z_i \\ +B \re \frac{d (z_i^2 z_i^*)}{dt} -\Omega^2 \re z_i
  -\gamma \re \dot  z_i \Bigr) \quad ,
\end{multline}
where the terms containing $\dot z_i$ can be substituted from
Eq.~(\ref{eq_zdot}).

These equations constitute a system of globally coupled oscillators
\cite{Pikovsky2001,golomb92,hakim92,nakagawa94} (the oscillators
interact with each other via the single variable $F$).  However, our
model differs in a crucial way from classical models of
synchronization, such as the Kuramoto model \cite{Kuramoto84}, in that
the coupling variable $F$ is determined by a first-order differential
equation (Eq.~\ref{eq_fdot}), rather than as a function of the
variables $z_i$.  By expressing all frequencies in terms of $\omega$
and amplitudes in terms of $\sqrt{\omega/B}$, and assuming $N \gg 1$,
the number of model parameters is reduced to four: $\epsilon/\omega$,
$\Omega/\omega$, $\gamma/\omega$ and $\zeta/N M \omega$.

\begin{figure}[htbp]
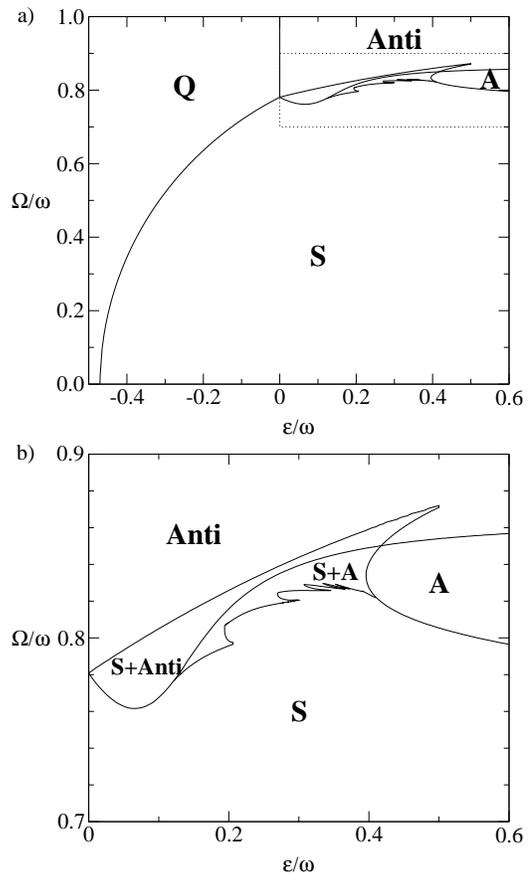

  \begin{center}
    \includegraphics{figure2a.eps}\\
    \includegraphics{figure2b.eps}
  \end{center}
  \caption{Phase diagram 
    (Q quiescent; S synchronous; Anti antisynchronous; A asynchronous) 
    as a function of the control parameter $\epsilon$
    of the active elements and the frequency $\Omega$ of the inertial
    oscillator.  Other parameters have the values
    $\zeta/NM\omega=0.3125$ and $\gamma/\omega=0.2$.  The diagram was
    established by numerical solution of Eqs.~(\ref{eq_zdot}) \&
    (\ref{eq_fdot}). The phase boundary of the synchronous state is
    given by Eq.~(\ref{eq:Omega}) for $\epsilon < 0$ and can be
    determined by perturbation theory for $\epsilon > 0$.}
\label{fig:phasediag}
\end{figure}

We start our analysis by considering the situation in which the active
elements would be stable if isolated, $\epsilon < 0$. With $\epsilon$
fixed, the frequency of the inertial oscillator $\Omega$ acts as a
control parameter for the system as a whole.  We can identify a
transition between a \emph{quiescent} phase (for $\Omega<\Omega_c$) in
which the entire system is at rest, and a \emph{synchronized} phase
(for $\Omega>\Omega_c$) in which all of the contractile elements and
the massive load oscillate together (see Fig.~\ref{fig:phasediag}). At
the critical value $\Omega=\Omega_c$ the system undergoes a Hopf
bifurcation. In its vicinity we can use linear stability analysis,
because the amplitude tends to zero there.  In the synchronized phase,
all oscillators have identical displacement $z_i \equiv z = x + {\rm
  i}y$ and Eqs.~(\ref{eq_zdot}) \& (\ref{eq_fdot}) form a system of
three coupled differential equations for $x$, $y$ and $F$. The
characteristic equation for the eigenvalues of the Jacobian reads
$\lambda^3 -(\epsilon
-\zeta/NM-\gamma)\lambda^2+(\Omega^2-\epsilon\gamma -2\epsilon
\zeta/NM)\lambda -\Omega^2\epsilon +(\epsilon^2+\omega^2)\zeta/NM=0$.
The transition occurs when the real part of the complex eigenvalue
pair changes sign, giving
\begin{equation}
\label{eq:Omega}
\Omega_c=\sqrt{\omega^2/\left(1+ \frac {NM\gamma} \zeta\right) +\epsilon\left( \gamma
    - \epsilon +2 \frac \zeta {NM} \right)} \quad .
\end{equation}
The frequency $f$ of synchronized
oscillations at the bifurcation is given by the imaginary part of the complex
eigenvalues:
\begin{equation}
f=\sqrt{
\omega^2/\left(1+\gamma NM / \zeta \right) -\epsilon^2 } \quad .
\end{equation}
Note that it is always the case that $\Omega_c<f<\omega$.  Away from
the bifurcation, $\Omega<\Omega_c$, the synchronized oscillations have
a frequency lower than $f$, thus $f$ represents the maximal frequency
of the system.

To interpret this result in terms of the underlying physical model, we
can consider two different ways of adjusting the load. In the first,
illustrated in Fig.~\ref{fig:phasediag}a, we fix the mass and the
damping of the inertial oscillator (thus $M=\text{const}$,
$\gamma=\text{const}$) and change the stiffness of the spring (a
stiffer spring implies a higher frequency $\Omega$ and vice versa).
Then a sufficiently strong spring, such that $\Omega > \Omega_c$, will
always maintain the stability of the system. But decreasing the
stiffness can provoke a transition to synchronized oscillations,
provided that the control parameter $\epsilon$ lies above some
threshold value, determined by the solution $\Omega_c=0$ of
Eq.~(\ref{eq:Omega}) (i.e.~the active elements must be sufficiently
close to their dynamical instability).  We also note that the chain
can oscillate synchronously even when the inertial oscillator itself
is overdamped (i.e.~$Q=\Omega/\gamma<1$).  In the second situation, we
vary the mass $M$ of the inertial oscillator while keeping the spring
stiffness $M\Omega^2$ and the damping $M\gamma$ constant. Now we find
that the system will remain quiescent if the mass is sufficiently
small, but that a larger mass can make the system oscillate provided
that $\epsilon > -\omega/\sqrt{1+ NM \gamma /\zeta}$.

\begin{figure}[t]
  \begin{center}
    \includegraphics{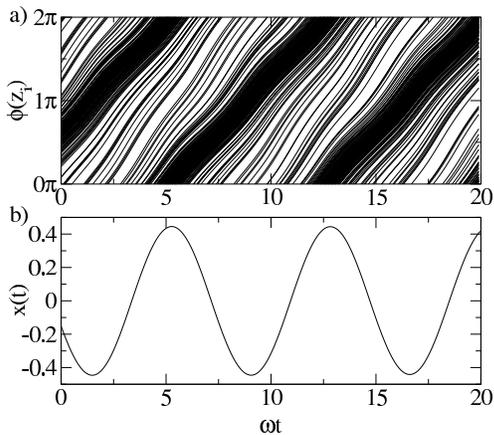}
  \end{center}
  \caption{The asynchronous state.  (a) Phase $\phi_i=\arg(z_i)$ of
    each of the $N=128$ active elements as a function of time.  The
    phase velocity $\dot \phi_i$ of a particular element is slow for a
    number of periods of oscillation, then speeds up over a few
    periods, then slows again.  The average frequency of the active
    elements is therefore faster than the speed of propagation of the
    the high-density area in the phase space, which is the frequency
    of the inertial load. (b) Total extension $x=\Sigma_i \re z_i$ (in
    units $N\sqrt{{\omega}/{B}}$) as a function of time, illustrating
    the lower frequency of the inertial oscillator.  Parameters:
    $\zeta/NM\omega=0.3125$, $\gamma/\omega=0.2$,
    $\epsilon/\omega=0.6$ and $\Omega/\omega=0.83$.}
  \label{fig:asynchronous}
\end{figure}

We continue the analysis by considering the situation where the active
elements are individually unstable, $\epsilon > 0$, and would
oscillate spontaneously if isolated. In this case the system displays
a greater variety of phases, as indicated in Fig.~\ref{fig:phasediag}.
For small values of $\Omega$, there is a \emph{synchronous} phase as
described above.  For large values of $\Omega$ there is an
\emph{antisynchronous} phase: All of the active elements oscillate
with the same frequency $\omega$, but with a distribution of phases
$\phi_i=\arg(z_i)$ such that the sum of their extensions is always
zero; thus the massive load remains stationary. When the inertial
frequency $\Omega$ quite closely matches the characteristic frequency
$\omega$ of the active oscillators, there is a remarkable phase in
which the individual elements oscillate at one frequency, while the
total extension oscillates at a different, lower frequency. We label
this phase \emph{asynchronous}. As shown in
Fig.~\ref{fig:asynchronous}, active oscillators form clusters with
different phases $\phi_i$ and their phase velocities $\dot \phi_i$
periodically slow down and speed up. The number of oscillators in each
cluster generally decreases with increasing $\epsilon/\omega$ and can
be as small as one, which is the case in the example shown in
Fig.~\ref{fig:asynchronous}. The ensemble of oscillators resembles a
set of vehicles on a congested ring road, which repeatedly enter a
traffic jam.  Because vehicles enter the jam at its rear end and leave
at the front, the jam progresses more slowly than the average speed of
an individual vehicle. In our model, the total extension depends on
the phase of the majority of active oscillators, and the oscillation
frequency of the inertial load therefore corresponds to the speed of
propagation of the traffic jam.

\begin{figure}[t]
  \begin{center}
    \includegraphics{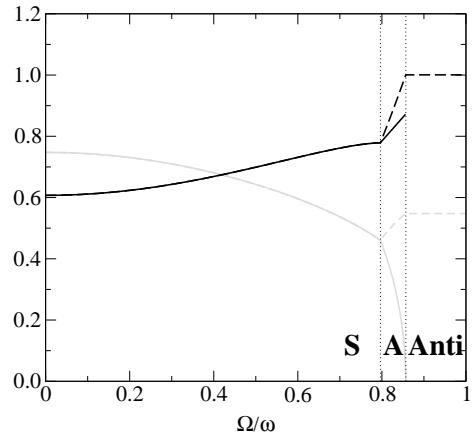}
  \end{center}
  \caption{Frequency and amplitude of oscillations as a function of
    $\Omega$ for a fixed, positive value of the control parameter
    ($\epsilon/\omega=0.6$). These parameters correspond to moving up
    the right vertical axis of Fig.~\ref{fig:phasediag}a. The system
    transits from the synchronous (S), to the asynchronous (A), to the
    antisynchronous (Anti) phase. Black curves show the frequencies of
    the inertial load (continuous) and of the active elements
    (dashed). Gray curves show the RMS amplitude of the mass
    (continuous) in units $N\sqrt{{\omega}/{B}}$ and that of each
    active element (dashed) in units $\sqrt{{\omega}/{B}}$. Other
    parameters: $\zeta/NM\omega=0.3125$, $\gamma/\omega=0.2$.}
\label{fig:amp_freq} 
\end{figure}

The dependencies on $\Omega$ of the frequencies of individual
elements, and of the total displacement, are shown in
Fig.~\ref{fig:amp_freq}. Also shown are the corresponding amplitudes
of motion. In the synchronous phase, both frequencies are equal and
the amplitude of the inertial load is $N$ times greater than that of
each active element.  In the asynchronous phase, the amplitude of the
inertial oscillator falls towards zero, while that of the active
elements approaches $\sqrt{\epsilon/B}$ --- the amplitude of
spontaneous oscillations for an isolated element. At the same time,
the frequency of the active oscillators approaches their
characteristic frequency $\omega$. In the antisynchronous phase the
frequency and amplitude of the active elements remain unchanging at
these limiting values, and the inertial mass does not move. For large
values of $\epsilon$ all transitions are second order, and the system
passes directly from the synchronized, to the asynchronous, to the
antisychronous phase as $\Omega$ increases.  For small, positive
values of $\epsilon$ the transitions become first order and there is a
region in which the synchronous phase and the asynchronous phase
coexist as two metastable states. At intermediate values of
$\epsilon$, a complex asynchronous phase arises in which three or more
clusters of oscillators are synchronized within each cluster, but the
clusters are out of phase with each other (see
Fig.~\ref{fig:phasediag}b).

One experimental situation which appears to correspond to the regime
$\epsilon > 0$ is the spontaneous oscillatory contraction (SPOC) of
skeletal muscle in conditions of high ADP/ATP ratio.  Oscillations of
the length of a myofilament have been observed when its end was
attached to a flexible microneedle \cite{yasuda96}, and oscillations
in the length of individual sarcomeres have been seen when the total
length was held fixed \cite{Okamura.Ishiwata1988}. These correspond to
the synchronous and antisynchronous phases of our model. Additionally
a `metachronal' phase, in which contractile waves propagate along a
myofilament, has been observed \cite{Anazawa.Ishiwata1992}. This
cannot be explained by our model. It would require either chemical
signaling between adjacent sarcomeres, or a gradient in one or more of
the sarcomere properties (e.g.  the number of myosin molecules
interacting with the thin filament) as suggested by Smith and
Stephenson \cite{smith94}.

The mechanics of insect flight muscle has been investigated in detail
by Machin and Pringle \cite{Machin.Pringle1959,Machin.Pringle1960}.
They found that a sudden increase in fiber length caused a subsequent
rise in tension --- the phenomenon known as `delayed stretch
activation'. Consequently a muscle subjected to a sinusoidal change of
length produced net work \cite{Machin.Pringle1960}. They also observed
that a muscle could be made to oscillate by attaching it to an
inertial load \cite{Machin.Pringle1959}, provided that the damping was
not too great.  They suggested that the load must be resonant
(i.e.~$Q=\Omega/\gamma>1$) for oscillations to occur, and found that
the frequency of vibration was primarily determined by the inertia and
elasticity of the load. All of these results are consistent with our
model in the regime $\epsilon < 0$ and $\zeta /N M \omega \ll 1$.  But
we predict that it is not in general necessary for the load to be
resonant; oscillations should be observable whenever $\Omega$ is
smaller than $\Omega_c$, given by Eq. (\ref{eq:Omega}). And we note
that the oscillation frequency generally depends on the characteristic
frequency $\omega$ of the sarcomeres, as well as on the nature of the
inertial load. It is often stated that the wings and thorax of insects
provide a resonant load whose oscillation is maintained by energy
supplied by the flight muscles
\cite{Dickinson97,Josephson.Stokes2000}. In the light of our
investigation, it would be interesting to conduct experiments in which
the mass and damping of this load is modified, to verify whether the
muscles can generate oscillations in the absence of resonance.
\begin{acknowledgments}
  We have benefited from discussions with Martin Falcke and thank
  Karsten Kruse for a critical reading of the manuscript.  A.V. would
  like to acknowledge support from the European Union through a Marie
  Curie Fellowship (No.~HPMFCT-2000-00522) and from the Slovenian
  Office of Science (Grant No.~Z1-4509-0106-02).
\end{acknowledgments}


\begin{thebibliography}{27}
\bibitem[{not()}]{note_myogenic}
\bibinfo{note}{{These flight muscles are variously called `myogenic',
  `fibrillar', or `asynchronous'; this last term refers to the fact that
  contraction is not directly related to the nerve impulses and has nothing to
  do with the synchronization of sarcomere oscillations discussed in this
  Letter.}}
\end{thebibliography}
\end{document}